# Modeling of dual-metal Schottky contacts based silicon micro and nano wire solar cells

M. Golam Rabbani[1], Amit Verma[2], Michael M. Adachi[3], Jency P. Sundararajan[1], Mahmoud M. Khader[4], Reza Nekovei[2] and M. P. Anantram[1]

[1]Department of Electrical Engineering, University of Washington, Seattle, WA 98195 USA,

[2]Department of Electrical Engineering and Computer Science, Texas A&M University – Kingsville, Kingsville, Texas 78363, USA,

[3]Department of Electrical and Computer Engineering, University of Toronto, Toronto, Ontario M5S 3G4, Canada,

[4]Gas Processing Center, College of Engineering, Qatar University, Doha, P.O. 2713, Qatar.

**Abstract:** We study solar cell properties of single silicon wires connected at their ends to two dissimilar metals of different work functions. Effects of wire dimensions, the work functions of the metals, and minority carrier lifetimes on short circuit current as well as open circuit voltage are studied. The most efficient photovoltaic behavior is found to occur when one metal makes a Schottky contact with the wire, and the other makes an Ohmic contact. As wire length increases, both short circuit current and open circuit voltage increase before saturation occurs. Depending on the work function difference between the metals and the wire dimensions, the saturation length increases by approximately an order of magnitude with a two order magnitude increase in minority carrier length. However current per surface area exposed to light is found to decrease rapidly with increase in length. The use of a multi-contact interdigitated design for long wires is investigated to increase the photovoltaic response of the devices.

1. Introduction

One dimensional nanomaterials like nanowires and nanotubes hold great potential for many applications such as electronics [1,2], sensors [3,4], and photovoltaics [5–7]. Nano engineered materials like nanowires and nanotubes are considered to be potential candidates for low cost and high efficiency solar cells. There have been many studies on solar cells based on single as well as multiple nanowires [8–11]. Tsakalakos *et al.* [8] studied p-n junction based silicon nanowire solar cells on metal foils, and found large current density and low optical reflectance. Sivakov *et al.* [9] fabricated silicon nanowire solar cells by electroless wet chemical etching of micro crystalline silicon layer on glass and achieved a high power conversion efficiency of 4.4%. Tian et al. [10] studied single p-i-n coaxial silicon nanowires and measured open circuit voltage ($V_{oc}$) of 0.26 V and short circuit current ($I_{sc}$) of 0.503 nA. Experimental study on Schottky solar cells comprising multiple SiNWs bridging two different metals with different work functions was carried out by Kim et al. [11]. They obtained a low $V_{oc}$ of 0.167 V but high $I_{sc}$ of 91.1 nA. Kelzenberg et al. [12] studied single-nanowire solar cells with one rectifying junction created by electrical heating of the segment of the nanowire beneath it. For a nanowire of diameter 900 nm, they achieved a $V_{oc}$ of 0.19 V and a short circuit current density of $5.0\,\mathrm{mA\,cm^{-2}}$. Hybrid Schottky diode solar cells [13] with poly(3,4-ethylenedioxythiophene) poly(styrenesulfonate) (PEDOT:PSS) film deposited on metal-assisted chemically etched SiNW arrays produced Voc of ~0.48 V and $J_{sc}$ of ~30 mA/cm². These works on nanowire based solar cells primarily focus on experimental investigations to demonstrate their potential in realizing the next generation of solar cells. However, a detailed study on the influence of various parameters like nanowire dimensions and work function of the metal contacts in modifying the photovoltaic behavior of the nanowires is lacking. In this work, we present results from simulation studies on Schottky junction based microwire and nanowire solar cells, and investigate the dependence of their photovoltaic properties on metal work functions, wire dimensions as well as minority carrier lifetimes.

**2. Device structure, problem statement**

Fig. 1 is a sketch of the device structure under study. There are two dissimilar metal pads, with dissimilar work functions, bridged by a rectangular cross-section wire. L, W and H represent the wire length, width and height, respectively. The array of downward pointing arrows represents vertically downward incident light beam. Only the top surface (of area L x W) of the wire is illuminated. The effect of substrate is not considered in this work. Performance enhancing features such as an antireflection coating or back reflector have not been included so as to keep the focus on the role of the silicon wire and metal contacts.

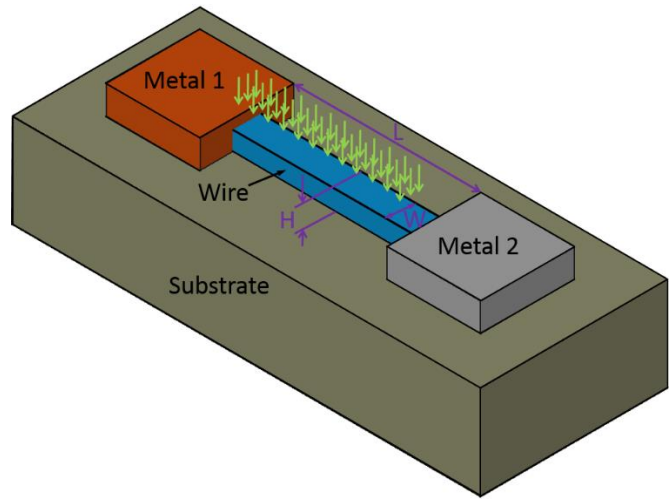

**Fig. 1.** Schematic representation of a single nanowire between two metal pads. Wire dimensions are indicated, and the downward pointing array of green arrows represents the incident light.

The work focuses on understanding how device dimensions and minority carrier lifetimes affect the photovoltaic properties (short circuit current, open circuit voltage, I-V characteristics) of the solar cell. The work also investigates the effects of the metal work functions on the solar cell performance. It explores ways to improving the efficiency of the solar cells, including the selection of metals. Simulation is done with Silvaco Atlas software [14].

## 3. Simulation versus analytical solutions

We start with a simulation that can be verified against a one dimensional (1D) analytical solution. For this we assume that the contacts are Ohmic. For the simulation part, a microwire with L=6.8μm, W=1.0μm, H=0.85μm was 2D simulated using the approach described in the previous section. For the analytical solution, we consider the one dimensional (1D) minority carrier diffusion equation along wire length in the presence of an electric field [15]:

$$D_n \frac{\partial^2 \Delta n}{\partial x^2} + \mu_n \frac{\partial}{\partial x}\left(E[n_0 + \Delta n]\right) - \frac{\Delta n}{\tau_n} + G_L = 0 \qquad (1)$$

where $D_n$ is carrier diffusion coefficient, $\mu_n$ is carrier mobility, $E$ is the electric field, $n_0$ is equilibrium carrier density, $\Delta n$ is photo-generated (excess) electron density, $\tau_n$ is electron lifetime, and $G_L$ is photo-generation rate. Note $E$ constant along length for Ohmic contacts.

Then excess minority carrier concentration, $\Delta n$, is found by solving Eq. (1), which is a linear second order differential equation having a solution of the form

$$\Delta n = A e^{m_1 x} + B e^{m_2 x} + C \qquad (2)$$

where

$$m_{1,2} = -\frac{\mu_n E}{2D_n} \pm \sqrt{\left(\frac{\mu_n E}{2D_n}\right)^2 + \frac{1}{D_n \tau_n}}$$. The constants A, B and C are found from the boundary conditions, $\Delta n(x=0)=0$ at the left contact and $\Delta n(x=L)=0$ at the right contact, and are given by

$$A = C\left(\frac{1-e^{m_2 L_x}}{e^{m_2 L_x} - e^{m_1 L_x}}\right), C = G_L \tau_n \text{ and } B = -A - C. \tag{3}$$

Here $G_L$ along the 1D line for analytical calculation is extracted from two dimensional (2D) $G_L$ generated by Atlas simulator. Eq. (2) along with Eq. (3) represents the analytical expression of the excess minority carrier density. Analytical expression for current can be calculated by first finding the current densities as given below

$$\begin{aligned} J_n &= q\mu_n n E + q D_n \nabla n \\ J_p &= q\mu_p p E - q D_p \nabla n \quad [\text{as } \nabla p = \nabla n] \end{aligned} \tag{4}$$

where E is the electric field, $D_{n(p)} = kT\mu_{n(p)}/q$ is electron (hole) diffusion constant. Then total current, I, is found by multiplying the total current density by the cross sectional area, A, of the wire, or $I = A(J_n + J_p)$.

Fig. 2 presents the comparison between the results from simulation (line with symbols) and analytical calculation (solid line) for our microwire with a uniform p-type doping density of $10^{15}$ cm$^{-3}$. Fig. 2(a) compares excess electron (minority carrier) density for different applied biases between the contacts. The photo generated minority electron density is symmetric with respect to the contacts for zero bias, but the peak density shifts towards the positive (right) contact ( at position = 6.8μm ) as the bias is increased. Fig. 2(b) plots the total current with one sun illumination as a function of the bias voltage in the range 0.0-0.5V. The plots display an excellent match between analytical calculations (solid line) and simulation results (symbol).

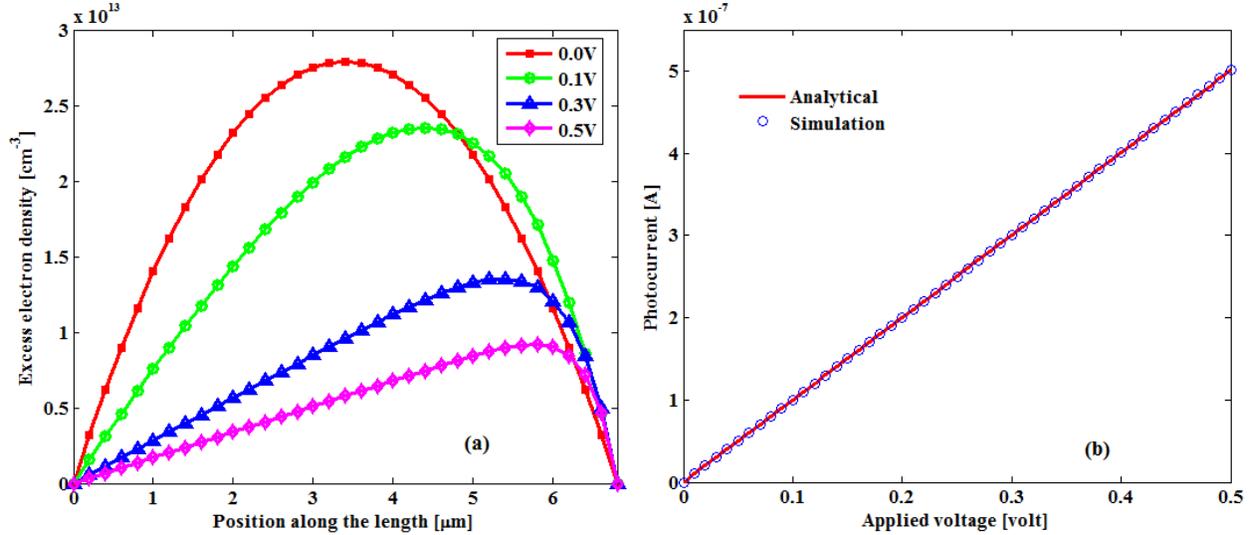

**Fig. 2.** Comparison of (a) excess electron density and (b) photocurrent from analytical calculation (solid line) and simulation (symbol).

The device operation can be understood as follows. At dark and equilibrium, hole density ($p$) throughout the device is equal to the doping density ($10^{15}$ cm$^{-3}$) while electron density is $n_i^2/p$ or $2\times10^5$ cm$^{-3}$. The current is obviously zero as there is no drift (due to zero electric field) or diffusion (due to zero density gradient). When a bias is applied at dark, due to Ohmic contacts, a constant electric field is set up along the length of the device. Hence the carrier densities are still the same as in the equilibrium case so that the diffusion currents are zero. However, drift currents proportional to the electric field (and carrier densities) are produced under bias. Since electron density is negligible compared to the hole density, the total current equals the hole drift current.

Under an illumination, the microwire and nanowire photovoltaic devices absorb photons of energy higher than the bandgap that excite electrons from the valence band to the conduction band. This process generates excess electron-hole pairs. Note however that the excess carrier density is zero at the contacts due to requirement of boundary condition. Since at short circuit (or zero applied bias) the device is symmetric with respect to the contacts, the excess carrier density is symmetric dome shaped in this case, as shown in the red curve of Fig. 2(a). This also implies a zero photo current at zero bias (zero drift) as excess carrier diffusions are equal and opposite. The excess carrier density is low for one sun illumination, so overall hole density is still dominated by the doping density. For electrons though the excess density is orders of magnitude higher than the equilibrium density, so Fig. 2(a) essentially plots the electron density under illumination. As the bias is increased, the applied electric field shifts the maximum excess carrier position towards the positive contact. In this case both drift and diffusion photo currents, and a nonzero total photo current, are produced. We find that the dark current for this device is too high because of the Ohmic contacts so that photocurrent to dark current ratio is close to unity.

Thus for Ohmic contacts, an asymmetric carrier profile due the applied bias causes nonzero (drift and diffusion) current as shown in Fig. 2(b). In addition to applied bias, asymmetry can also be introduced by dissimilar doping of the wire ends as well as by unequal work function Schottky contacts as discussed in section 5. For nanodevices, it is difficult to control the doping [16–19], while use of unequal work function metal contacts is relatively easier. Such an approach is also common in organic photovoltaics. Hence we study the effect of introducing asymmetry between the nanowire ends with dissimilar work functions on the photocurrent.

**4. Photocurrent versus wire length**

In all results below, standard solar spectrum air mass 1.5 (AM1.5) has been used whenever light is present. The recombination processes that we consider are Shockley-Read-Hall (SRH) and Auger recombination. Varying values of minority carrier lifetimes used in this work are assumed to include a range of bulk and surface recombination values.

Minority carrier lifetime is an important parameter for solar cell devices. Intuitively, the longer the wire, the more the surface area to absorb light and the larger will be the number of excess carriers generated inside for a single device and hence the larger the current. However, only carriers that reach the contact contribute to any photocurrent. The rest of the electron-hole pairs recombine inside the device. A longer lifetime gives minority carriers more time to travel to the contact before recombination with a majority carrier. On the other hand, if the lifetime is too short, excess minority carriers recombine with majority carriers before reaching the contact and as such will not contribute to the photocurrent. Thus carrier lifetime sets a limit to the maximum wire length beyond which photo generated carriers are not collected efficiently and as a result the photocurrent saturates. Hence the wire lengths and minority carrier lifetimes are related. This is seen in the results shown in Fig. 3, which plots zero bias photocurrent (short circuit current) of the wires of varying lengths for minority carrier lifetimes of 10μsec (solid curve), 1μsec (dashed curve) and 0.1μsec (dash-dotted curve). We considered doping density of ~$10^{15}$cm$^{-3}$ (p-type) for which minority carrier lifetime in crystalline bulk silicon is larger than 10μsec [20,21]. However, due to large surface to volume ratio, nanowires can have high a surface recombination rate that can potentially reduce minority carrier lifetime [22,23]. Life time that is an order of magnitude shorter than that of bulk silicon has been reported [23]. There have also been studies to improve the surface recombination and increase the lifetime [24,25]. In view of these results, the lifetime values we have selected are representative. Metal work functions of 5.5eV (left contact) and 4.0eV (right contact) have been used in this simulation.

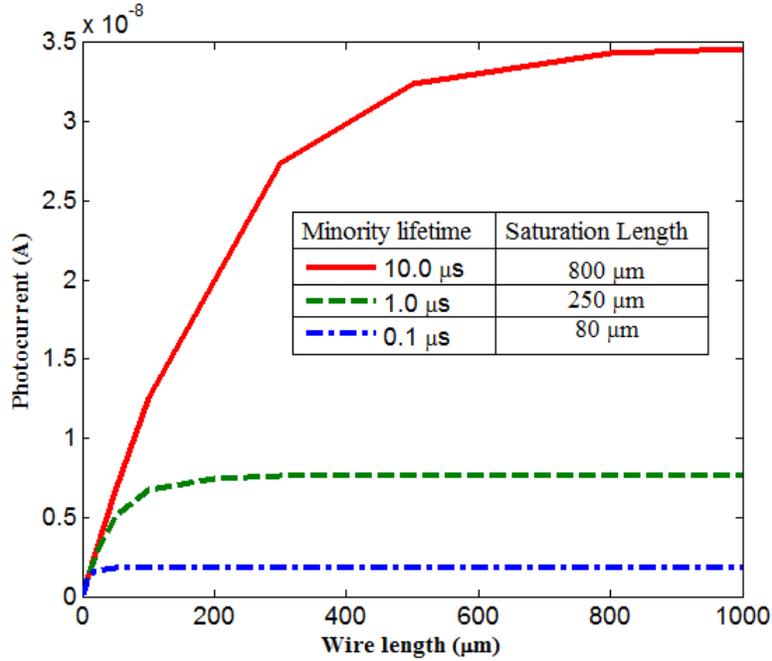

**Fig. 3.** Photocurrent *vs* wire length for three different minority carrier lifetimes.

Fig. 3 inset tabulates the relationship between the minority carrier lifetime and the device length at which the photocurrent approximately saturates. The saturation length is approximately proportional to the square root of the lifetime. This is reasonable given that the minority carrier diffusion length is defined as $L=\sqrt{D\tau}$, where $D=(kT/q)\mu$ is the diffusion constant, $\mu$ is the carrier mobility, and $\tau$ is the lifetime. At 300K, $L \simeq 50\mu m$ for $\mu=1000$ cm$^2$/V-sec and $\tau=1\mu sec$. For this case saturation length is about $250\mu m$. This relation holds for other lifetimes as well. Hence the saturation length is about five times the minority carrier diffusion length. This may be a good design parameter for nanowire based solar cell.

**5. Effect of metal work function**

As stated earlier, the results in Fig. 3 are for metal work function pair of 5.5eV and 4.0eV. This choice depends on the work function of usable metals and of course, the work function of the nanowire itself. The larger the difference between the two metal work functions, the larger the asymmetry, and the better the photovoltaic properties. However, in practice not all metal combinations may ideally be suitable. For example, both gold and platinum have high work functions, but they concomitantly also reduce the charge carrier lifetime [26,27] and thus may not be suitable as contact metals. On the low work function side, calcium and magnesium are highly reactive [28] and difficult to deposit since they oxidize very fast. Therefore for practical considerations, it is appropriate to study how lower work function differences affects solar cell behavior. Work function of the silicon nanowire can depend on many parameters such as etching time [29], chemical used in surface passivation, nanowire diameter [30], and doping. Silicon nanowire work function has been found to vary from about 4.5 eV to 5.01 eV [29,30]. Work function of silicon microwire is taken from the bulk silicon, which has an electron affinity of 4.17 eV. Therefore depending on doping type and concentration, the work function can vary from 4.17 eV (bottom of conduction band) to 5.25 eV (top of valence band) assuming a bandgap of 1.08 eV. In our study, we mostly considered the wire work function of 5.01 eV. This corresponds to bulk silicon with a p-type doping of $\sim 10^{15} cm^{-3}$. As explained through Fig. 4 below, the work function of one metal should be below and that of the other metal should be above that value. Thus considering a wire work function of 5.01 eV and work function range of usable metals, metal work functions of 5.5 eV and 4.0 eV are reasonable choices in our work.

Fig. 4 shows the conduction band (solid line) and the valence band (dashed line) as well as the Fermi level (dash dotted line) for a 10μm long wire. When there is no metal work function difference (Fig. 4(a)), the two contacts are identical. If light is incident uniformly on such a device both the photocurrent and the photovoltage will be zero due to the symmetry of the bands.

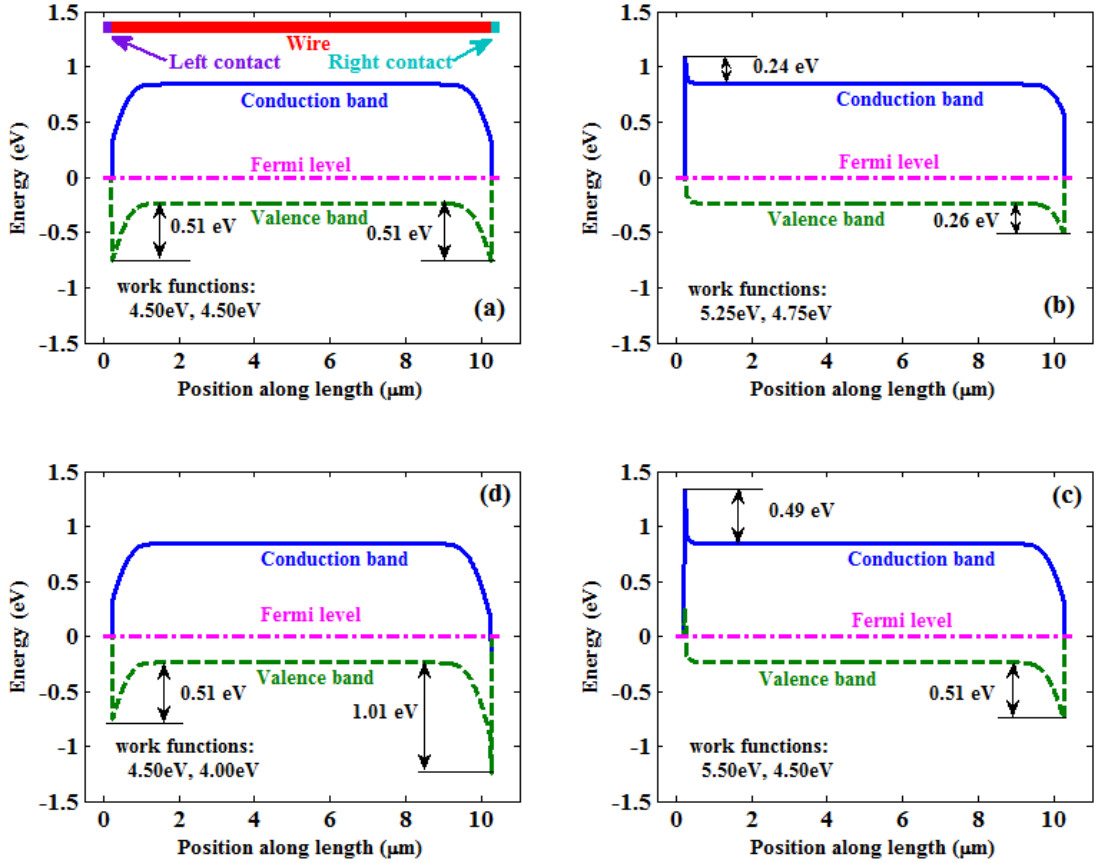

**Fig. 4.** Conduction (solid) and valence (dash) bands for a 10μm long wire. Contact work function pairs are 4.50eV, 4.50eV (a), 5.25eV, 4.75eV (b), 5.50eV, 4.50eV (c), 4.50eV, 4.00eV (d). Wire doping density is $10^{15} cm^{-3}$ (p-type, work function is 5.01eV). Inset of (a) at the top shows the device schematic, and labels the contacts and the wire.

Fig. 4(b) shows the case with work function of 5.25eV for left contact (larger than silicon work function) and 4.75eV right contact (less than silicon work function). Here electrons will prefer to flow towards the right contact while holes towards the left, resulting in a net photocurrent. If the barrier to the electron (hole) flow on the left (right) side is increased, photocurrent will also increase (Fig. 4(c)). This is described in more detail in relation to Fig. 8 below. If the work functions are chosen in such a way that bands at both ends bend in the same direction (Fig. 4(d)), relatively smaller photocurrent will be produced and the device will be inefficient. Section 5 discusses guideline for choosing the metal work functions for improved short circuit current and open circuit voltage. It is also to be noticed that although the band bending in Fig. 4 is affected by proper metal work function selection, controlled doping may also produce similar effects.

To see the effect of the work function difference on the photocurrent, we have plotted the short circuit photocurrent of the microwire device (width = 1μm and height = 35nm) as function of its length in Fig. 5(a). There are three curves for three metal contact work function differences. For a work function difference of 0.0 eV (dash dotted line), the current is zero, as can be expected from Fig. 4(a). For a moderate work function difference of 0.5 eV (dashed line), there is a considerable amount of photocurrent. This should be expected from the band diagram of Fig. 4(b). For a large work function difference of 1.5 eV (solid line), the current is even larger (see Fig. 4(c)), especially before saturation occurs at very long lengths. So the photocurrent increases with the work function difference.

Another aspect to consider is to compare the effect of dissimilar doping at the ends of a wire with the effect of work function difference of the metal pairs. Fig. 5(b) plots the short circuit photocurrent of the same wire as in Fig. 5(a) for three different doping concentrations ($10^{18}$cm$^{-3}$ (dash dotted line), $10^{19}$cm$^{-3}$ (dashed line), $10^{20}$cm$^{-3}$ (solid line)). Comparison of Fig. 5 (a) and (b) reveals that metals with work function difference of 1.5eV (work functions of 4.0eV and 5.5eV, solid line in Fig. 5(a)) can give the same short circuit photocurrent as that produced by doping concentrations equal to or in excess of $10^{19}$cm$^{-3}$ (Fig. 5(b), which makes the use of dissimilar metals an attractive alternative since controlled doping in nanowires is difficult [16–19].

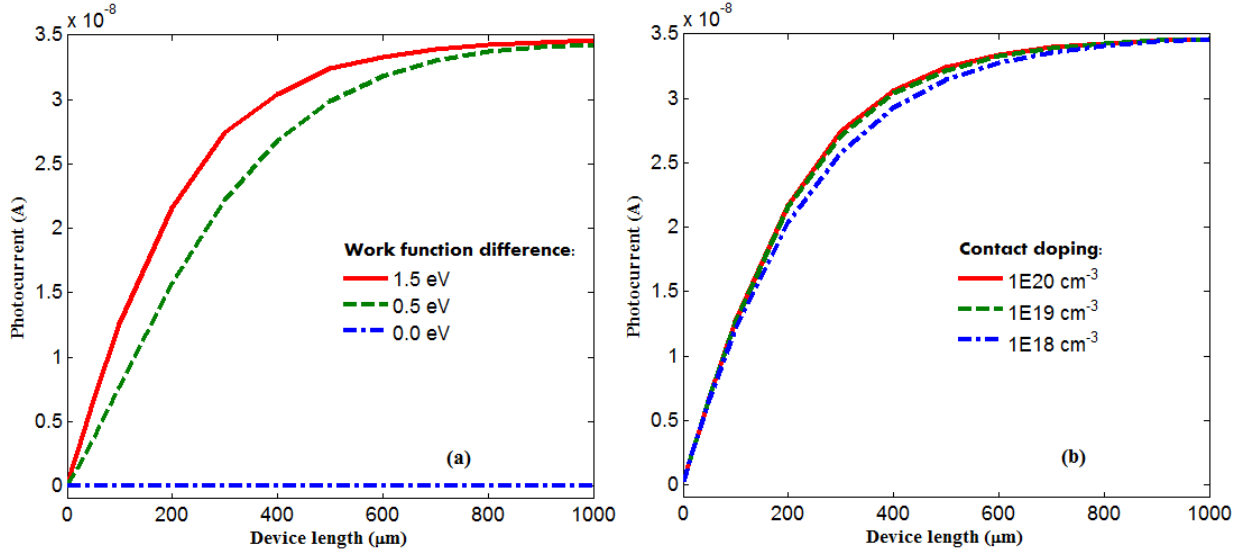

**Fig. 5.** Effect of (a) work function and (b) contact doping on short circuit photocurrent versus device wire length.

## 6. Short-circuit current and open-circuit voltage

Short-circuit current and open-circuit voltage are among the most important performance parameters of a solar cell. Short-circuit current ($I_{SC}$) is simply the photocurrent when the two electrodes are shorted to each other. Open-circuit voltage ($V_{OC}$) is the voltage across the device when the electrodes are open. In Fig. 6, we consider both $I_{SC}$ and $V_{OC}$ of wires of two different widths: 1 μm (a) and 100 nm (b). The lengths of the wires vary from 1 μm to 1000 μm. Each wire has a height of 35 nm. The metal work functions are 5.15eV (Nickel) and 4.15eV (Aluminum). Fig. 6(a) plots the 2D simulation results for the 1 μm wide wire. Note that 2D simulation is done on a vertical cross section of area L times H of the wire (Fig. 1). Since the height of the wire is still thin (35 nm), we also did the more computationally costly 3D simulation for this device (results not shown here). We have found that although the 2D and 3D results in our work have a close match, compared to 3D, the 2D simulations does tend to slightly overestimate both the $I_{SC}$ and $V_{OC}$. Thus for the narrower wires, only 3D (Fig. 6 (b)) simulations were done. $I_{SC}$ for the 1 μm wide wire is about 10 times that of the 100 nm wide wire. Indeed, from 3D simulation of the wider wire, it is found that the current scales exactly linearly with the wire width. There have been several simulation studies [31–34] on optical absorption of single and multiple silicon nanowires as function of their diameter. Most works are on arrays of vertical wires [34] with core-shell geometry [31]. Simulations of core-shell single cylindrical wire predict [31] that there is an optimal wire radius for maximum current density. It was also pointed out that simulation of single nanowire may not capture all the physics present in an array of wires. Current per unit area was found to increase while current per unit volume was found to decrease with wire diameter in a study [34] of horizontal hexagonal wires. Simulation study [32] of arrays of cylindrical wires lying horizontally on a flat surface with wire to wire distance of 200 nm predicts enhanced absorption with increasing diameter, ranging from 50 to 160 nm. Our result is consistent with that study. For our single rectangular wire, lying horizontally on a flat surface, the surface area over which normal light is incident upon it is exactly proportional to its width. Thus the photocurrent generated by absorption of a uniform intensity light is proportional to the wire width.

For both cases, the $I_{SC}$ first increases and then saturates with length. This is because as length increases, the probability of recombination of excess charge carriers also increases as they move towards their respective contacts. $V_{OC}$, as shown in the inset of Fig. 6, has a trend similar to that of $I_{SC}$. This length is 800 μm for both $I_{SC}$ and $V_{OC}$ for minority carrier lifetime of 10 μs.

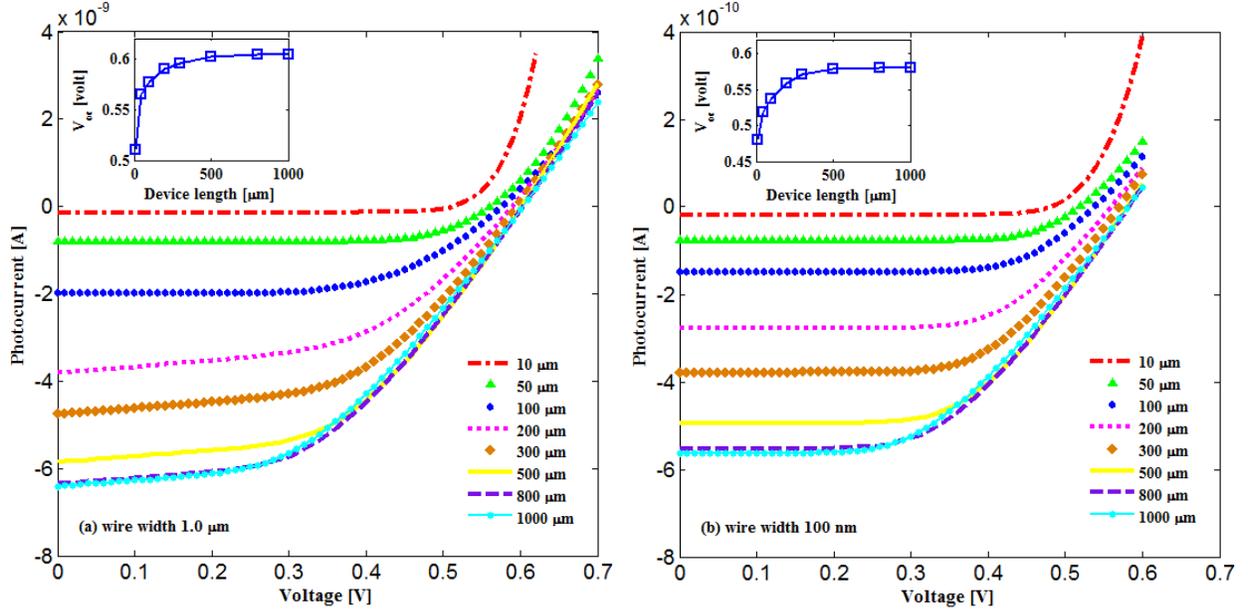

**Fig. 6.** Photocurrent versus bias for different nanowire lengths. (a) 2D simulation of 1 micrometer wide wire, (b) 3D simulation of 100 nanometer wide wire. Both wires have a height of 35 nanometer. Minority carrier lifetime is 10μs.

The behavior of Voc can be understood in terms of quasi-Fermi level splitting under illumination [35,36]:

$$V_{OC} = \frac{kT}{q}\ln\left(\frac{np}{n_i^2}\right) \simeq \frac{kT}{q}\ln\left(\frac{\Delta n(N_A + \Delta p)}{n_i^2}\right) \qquad (5)$$

where n (p) is the electron (hole) density, $N_A$ is the p-type doping density, $\Delta n$ ($\Delta p$) is the excess electron (hole) density due to illumination and $n_i$ is the intrinsic carrier density. Similar expression holds for n-type doping. If the device is very long such that in steady state the excess carrier densities approach the value $G_L \tau$, then Eq. (5) becomes,

$$V_{OC} \simeq \frac{kT}{q} \ln\left(\frac{(G_L\tau)(N_A+(G_L\tau))}{n_i^2}\right) \qquad (6)$$

Eq. (6) gives a limit on the available Voc. In an actual device, the closer the average excess densities are to $G_L\tau$, the closer its $V_{OC}$ will reach the value given by Eq. (6). Fig. 7 plots available as well as actual Voc as a function of the AM1.5 light intensity for different wire lengths and two different work function differences between the Schottky contact and the wire.

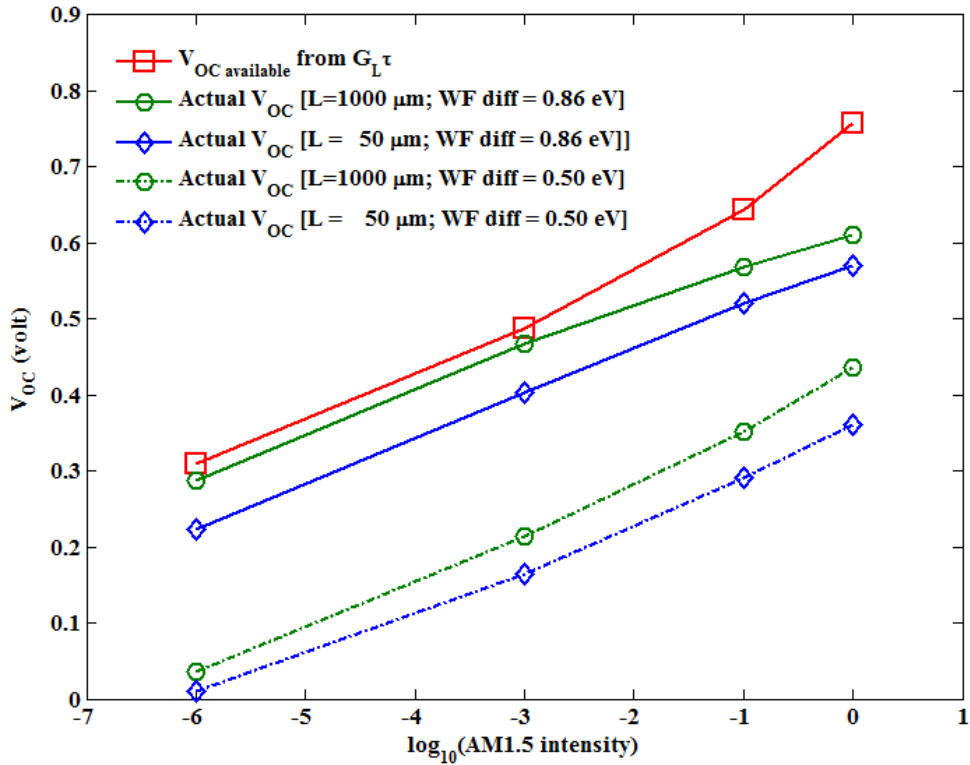

**Fig. 7.** Available and actual open circuit voltages as a function of intensity. Wire doping density is $10^{15} cm^{-3}$ (p-type, work function is 5.01eV). Lifetime is $10\mu s$.

Long wire with large work function difference approach the ideal device as can be seen for the case of 1000μm long wire with work function difference of 0.86eV. For shorter wire (50μm), with the same work function difference, the voltage is lower. The reason is that the shorter the wire the lower its excess carrier density is compared to $G_L \tau$.

Since appropriate contact work function values are important for the nanowire solar cells, we discuss below how one may select the work functions to get larger $I_{SC}$ and $V_{OC}$. The wire that we consider is 100μm long, 1μm wide and 35nm thick. The p-type doping density of the wire is $10^{15} cm^{-3}$ so that the Fermi level is fixed at 5.01 eV below the vacuum level (conduction band is at 4.17 eV and valence band is at 5.25 eV below). As discussed in Fig. 4, in this case, work function of the Ohmic contact ($W_O$) should be below and the work function of the Schottky contact ($W_S$) should be above, the wire equilibrium Fermi level.

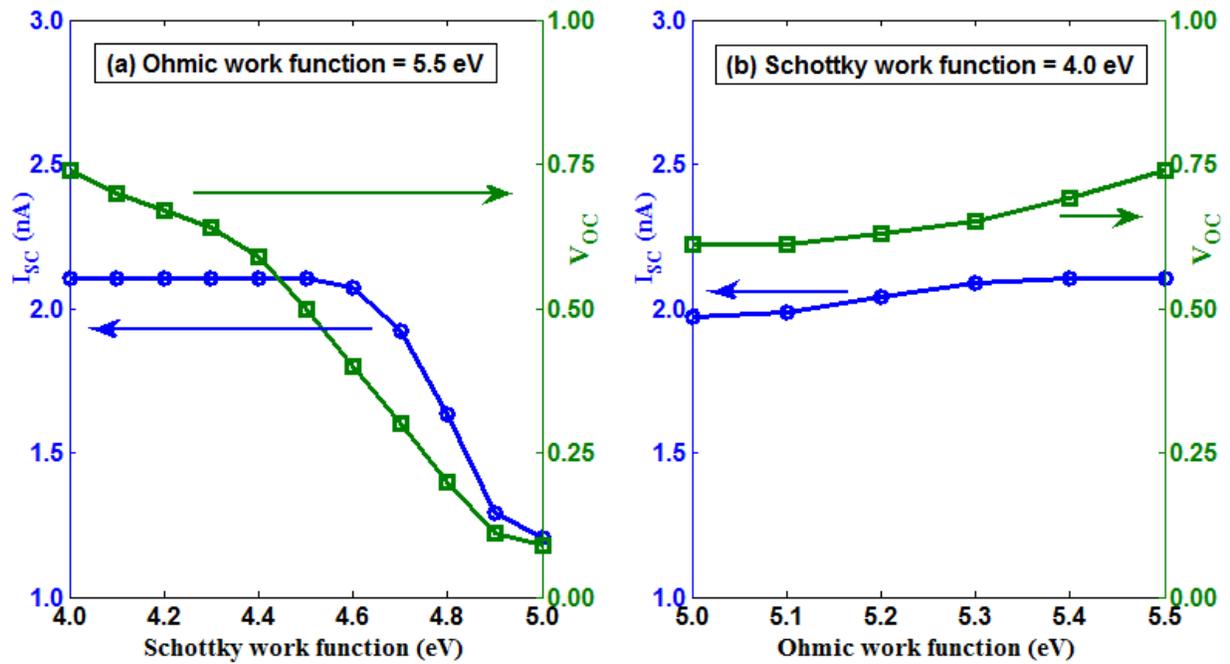

**Fig. 8.** $I_{SC}$ and $V_{OC}$ as a function of (a) $W_S$, with $W_O$ fixed, (b) $W_O$, with $W_S$ fixed.

Fig. 8(a) plots $I_{SC}$ and $V_{OC}$ as a function of the $W_S$ with $W_O$ kept fixed at 5.5 eV while in Fig. 8(b) the $W_S$ is fixed at 4.0 eV and $W_O$ changes. As $W_S$ moves above and away from the wire work function (Fig. 8(a)), both $I_{SC}$ and $V_{OC}$ increase rapidly. But $I_{SC}$ saturates when $W_S$ is about 0.5 eV above the wire work function, while $V_{OC}$ keeps increasing at a slower rate. On the other hand, variation in $W_O$ (with a fixed $W_S$) does not produce as high a change in both the open circuit voltage and short circuit current (Fig. 8 (b)). We find that $I_{SC}$ saturates when $W_O$ is 0.4 eV below the wire work function. Thus for efficient solar cell the work function of the Schottky contact is very important and should be as far away from the wire's work function as possible. The work function of the Ohmic contact although less important, can still cause a noticeable change in the $V_{OC}$, which should be kept in mind when selecting the metal.

## 7. Short-circuit current density

We have so far considered the total currents of the individual devices (microwire and nanowire) without regard for their surface areas. Since the amount of device area exposed to light is very important for solar cells, we define a short circuit current density as

$$J_{SC} = \frac{I_{SC}}{LW} \quad (7)$$

where $I_{SC}$ is the total short circuit current and $J_{SC}$ is the short circuit current density. L and W are wire length and width, respectively. Note that this definition is different from short circuit current density used in conventional planar solar cells where the current flows perpendicular to the surface area. For the horizontally lying wires we considered in this study, the area in Eq. (7) is the area exposed to light (Fig. 1), and the current flows parallel to this area.

We plot the short circuit current density ($J_{SC}$) for both the microwire and the nanowire devices for three different representative minority carrier lifetimes in Fig. 9. In this plot, surface areas of the two metal contacts at the ends of the wire have not been taken into account. For a particular minority carrier lifetime, $J_{SC}$ for both microwire and nanowire are the same, and thus $I_{SC}$ scales with wire surface area, as discussed in relation to Fig. 6. $J_{SC}$ decreases as wire length increases, and the shorter the lifetime the faster the decrease. The reason is again attributed to the increased inefficient collection of photogenerated carriers as the length increases.

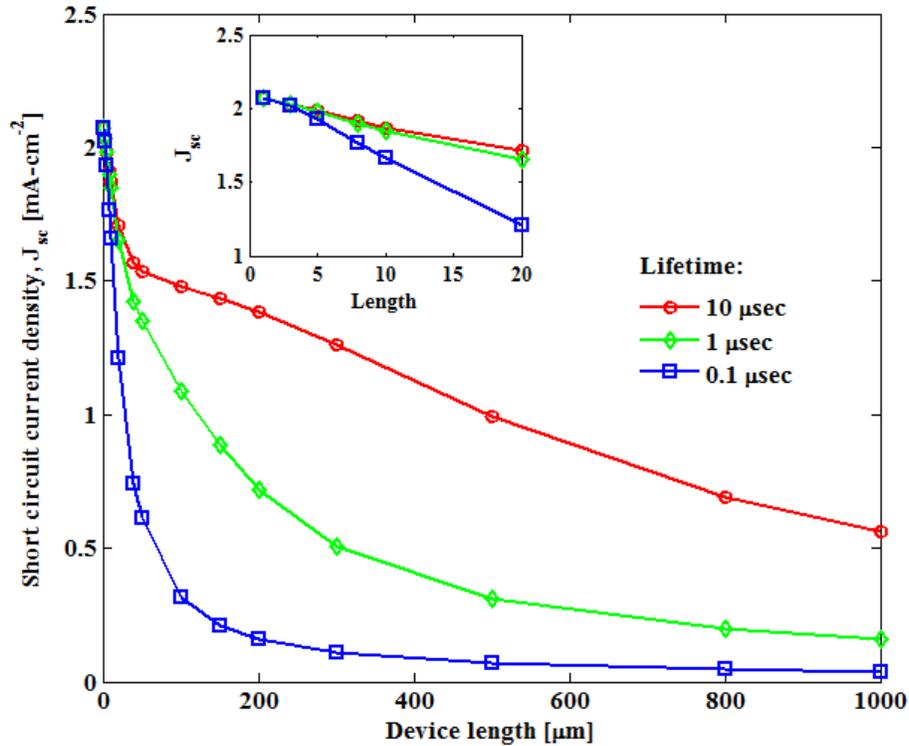

**Fig. 9.** Short circuit current density versus wire lengths for three different minority carrier lifetimes. The wire has a height, H, of 35 nm.

As seen in Fig. 3, the $I_{SC}$ for a wire saturates at 5 times the diffusion length. However, when the surface area is taken into account, the carrier collection efficiency decreases rapidly with length. This implies that to obtain large current density per surface area, shorter wires are preferable. However for greater current per nanowire, longer wires are preferable.

**8. Comparison with reported experimental results**

Kelzenberg et al. [12] measured both $V_{OC}$ and $J_{SC}$ for single silicon nanowires of diameter 900 nm and length 20μm and it is interesting to compare our predictions with their measurements. However, there is one important difference. The reported work uses aluminum for both contacts, where one contact is electrically heated to get Schottky effect. Our work considers specific work functions corresponding to two distinct metals. We simulate a wire of the same dimension with a minority carrier lifetime of 15nsec, as given in [12]. For aluminum, the work function is approximately 4.15 eV. However, the work function of heated aluminum-silicon interface is not well understood [12] and the exact value is not available. Here a value of 4.52 eV gives a good match with experiment. We find $V_{OC}$ of 0.193V and $J_{SC}$ of 4.2 mA/cm$^2$, which are comparable to 0.19V and 5 mA/cm$^2$, respectively.

## 9. Improving Short Circuit Current

For long wires, saturation of short circuit current is a drawback. However, it is possible to make an improvement with modifications. So far we have considered contacts only at the ends of the wire. Fig. 10 depicts a sketch of a wire with additional metal contacts placed between the two ends. This contact arrangement essentially breaks a long wire into a few short nanowires electrically, with the cathodes connected together and the anodes connected together. This causes collection efficiency to increase since electrons and holes have to travel shorter distances.

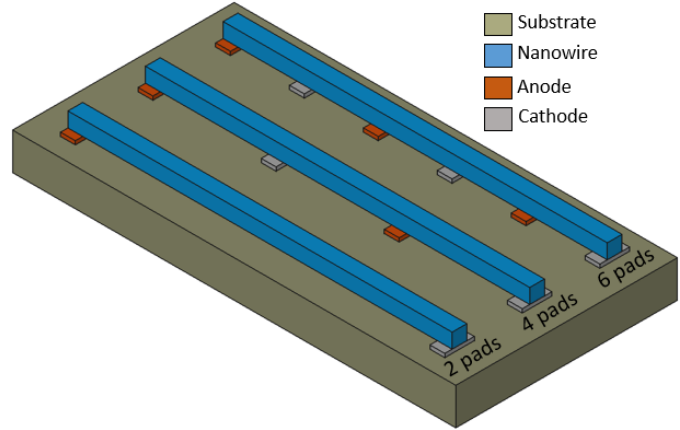

**Fig. 10.** Electrically breaking a long wire into a few short wires.

Fig. 11 plots both $I_{SC}$ (a) and $V_{OC}$ (b) versus the number of pads for nanowires of 6 different lengths. Here a minority carrier lifetime of 10μs has been considered. For lengths below 220μm, more than two pads decrease the current (Fig. 11(b)). But for lengths 220μm and above, maximum current is obtained when more than two pads are used. The results in Fig. 11(a) potentially imply that for wires with shorter lifetimes (more defects) employing multiple pads will improve collection efficiency even at shorter lengths. In addition, there is an improvement in fill factor (not shown) when number of pads equal or exceed the number required for maximum current. The $V_{OC}$ vs number of pads plot in Fig. 11(b) indicates that for wire length of 500μm or larger, open circuit voltage is greater for more than two pads. For number of pads maximizing the short circuit current (Fig. 11(b)), $V_{OC}$ is still larger than its value with two pads, if length is 500μm or larger.

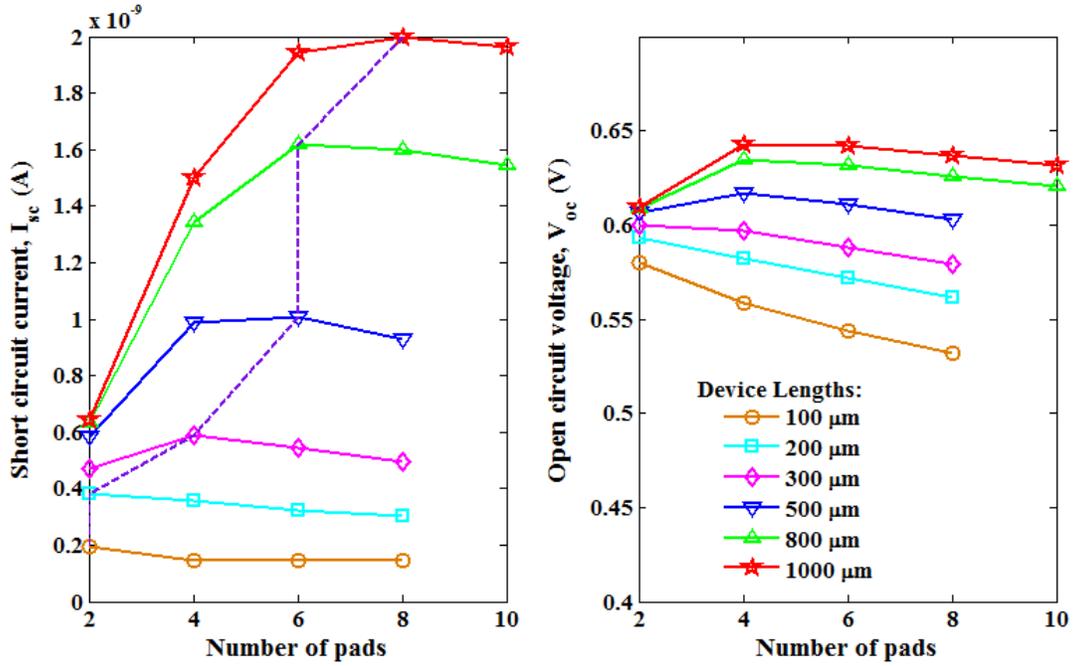

**Fig. 11.** Short circuit current (a), and open circuit voltage (b) as a function of number of pads for six different nanowire lengths. The wire widths and heights are fixed at 100nm and 35nm, respectively. Minority carrier lifetime is $10\mu s$.

Electrically connecting silicon wires in series or parallel can be done using current technology. A long silicon wire that is electrically broken by an interdigitated electrode pattern, as discussed above, may be thought of as a series connection of many smaller wires. The fabrication of millimeter long silicon nanowires has been reported some time back [37–40]. Electrode pattering for the long wires follows well defined fabrication procedures. On the other hand, arrays of silicon nanowires between two electrodes can be viewed as a parallel connection of those nanowires. Such arrays (or mats) have demonstrated optical properties that make them attractive candidates for solar cells, such as enhanced optical absorption [41].

**10. Conclusion**

Nano and microwire photovoltaic devices have been investigated. The use of two dissimilar metals contacts, with dissimilar work functions, for these devices is seen to be an excellent alternative to dissimilar doping at the two ends of the wire. This implies that issues related to fabrication reliability of such devices can be significantly reduced because controlled doping in nanowires is difficult and expensive. The behavior of nanowire photovoltaic devices is found to be dependent on the wire length for short wires. However for longer wires, both the short circuit current and open circuit voltage saturate. This saturation length is found to be approximately five times the minority charge carrier diffusion length, and suggests an upper bound on the achievable photocurrent. This bound can be overcome through modifications of the basic structure. Use of interdigitated patterns of dissimilar metals for a very long wire is seen to significantly increase the short circuit current, while keeping the open circuit voltage nearly constant. This is attributed to increased collection of charge carriers before they recombine.

**Acknowledgements** The work of M. Golam Rabbani and M. P. Anantram was supported by the National Science Foundation under Grant No. 1001174. Jency Sundararajan and M. P. Anantram were also partially supported by the QNRF grant (NPRP 5 – 968 – 2 – 403) and the University of Washington. Amit Verma, Mahmoud M. Khader, and Reza Nekovei were supported by the QNRF grant (NPRP 5 – 968 – 2 – 403).